\documentclass[prl,aps,twocolumn,superscriptaddress]{revtex4-2}
\usepackage[T1]{fontenc}
\usepackage{amsmath,amssymb}
\usepackage{amsfonts}
\usepackage{epsfig,pstricks,graphicx}
\usepackage{bm}
\usepackage{physics}
\definecolor{armygreen}{rgb}{0.29, 0.33, 0.13}
\usepackage{color}
\usepackage{array}
\newcommand{\av}[1]{\left \langle #1 \right\rangle}

\usepackage[normalem]{ulem} 
\usepackage{soul} 
\newcommand\reduline{%
 \bgroup\markoverwith
  {\textcolor{yellow}{\pgfsetfillopacity{0.2}\rule[-0.5ex]{2pt}{10pt}\pgfsetfillopacity{1}}%
   \textcolor{yellow}{\llap{\rule[0.4ex]{2pt}{0.4pt}}\llap{\rule[0.7ex]{2pt}{0.4pt}}}%
  }%
  \ULon}
\usepackage{lineno}

\begin{document}

\title{
 On detecting violation of local realism with photon-number resolving weak-field homodyne measurements}
 
\author{Tamoghna Das}
\author{Marcin Karczewski}
\author{Antonio Mandarino}
\author{Marcin Markiewicz}
\author{Bianka Woloncewicz}
\author{Marek \.Zukowski}
\affiliation{International Centre for Theory of Quantum Technologies, University of Gda\'nsk, 80-308 Gda\'nsk, Poland}

\begin{abstract} 
Non-existence of a local hidden variables (LHV) model for a phenomenon  benchmarks  its use in device-independent quantum protocols. Nowadays photon-number resolving weak-field homodyne measurements allow realization of emblematic gedanken experiments. Alas, claims that we can have no LHV models for such experiments on (a)  excitation of a pair of spatial modes by a single photon, and (b) two spatial modes in a weakly squeezed vacuum state, involving {\it  constant local oscillator strengths,} are unfounded.
For (a) an exact LHV model  resolves the dispute on the ``non-locality of a single photon'' in its original formulation. It is measurements  with {\em local oscillators on or off} that do not have LHV models.

\end{abstract}

\maketitle
Recent trailblazing experiments \cite{WEAKHOMODYNING-2009, WEAKHOMO-OPTICAL-COHERENCE,WALMSLEY}, involving weak-field homodyne detection with photon-number resolution  \cite{WEAKHOMODYNE-THEORY}, demonstrate that optical phenomena which exhibit both particle and wave nature of light are now possible to observe.
Gedankenexperiments once more turn real.
For instance,  Ref.  \cite{WEAKHOMO-OPTICAL-COHERENCE} reports photon-number resolving weak-field homodyne measurements in the case of the two emblematic signal field entangled states of two spatial optical modes, namely:

- (a)  excitation of a pair of spatial modes by a single photon (first discussed in \cite{TWC91}),

- (b) two spatial modes in a squeezed vacuum state (first suggested in \cite{GRANGIER-P-Y}).

In Ref. \cite{WEAKHOMO-OPTICAL-COHERENCE} the detectors were able  to distinguish photon numbers with a significant probability. State of the art techniques  allow now this to up to 20 photons with approx 90\% efficiency \cite{WALMSLEY}. 

The authors of \cite{WEAKHOMO-OPTICAL-COHERENCE} express the hope that the setups, when 
perfected, could lead toward ``the quantitative study of the non-local properties of multimode states'', including violation of local realism.
This suggests, that such operational arrangements could be used for device-independent implementations of quantum information protocols, e.g. key distribution, or random number generation. 

The trait  of experiments (a) and (b), and their realization in Ref. \cite{WEAKHOMO-OPTICAL-COHERENCE} is that they use at the measurement stage constant local oscillator strengths and 50-50 beamsplitters.
We construct an explicit LHV model for the ideal predictions of 
experiment (a), and show that the hope for a violation of local realism in experiment (b), in \cite{GRANGIER-P-Y} and \cite{WEAKHOMO-OPTICAL-COHERENCE} is not substantiated (a partial LHV model exists, and claims about violation of Bell inequalities are at least premature).
On the positive note we show that 
Bell-type experiments on the signal states (a) and (b) with each observer switching the local oscillator on (setting 1) or off (setting 2), and involving (non 50-50) optimized beamsplitters,  do violate a Bell inequality.  Thus we show a proper operational scenario for Bell experiment involving weak-field homodyne measurements in the case of both experiments.

This signals once more that a great care must be taken when claiming Bell non-classicality (for earlier controversies of this type involving other experiments see e.g. \cite{Mandel88} and \cite{PSHZ97}, \cite{Franson89} and \cite{Franson99}).

Our explicit precise model closes the long standing  dispute on whether {\em the} 1991 gedankenexperiment  by Tan, Walls and Collett (TWC) \cite{TWC91} reveals ``nonlocality of a single photon''. Moreover, we show that thus far \emph{there is no evidence} that the 1988 proposal of Grangier, Potasek and Yurke (GPY) \cite{GRANGIER-P-Y} involving parametric down conversion (intuitively an emblematic source of entanglement) and weak-field homodyning  (with constant local oscillator strengths, and 50-50 beamsplitters) constitutes a valid Bell-type experiment. We conjecture that a full local realistic model for it exists.

\begin{figure}
	\centering
\includegraphics[width= 1. \columnwidth, height= 0.7 \columnwidth]{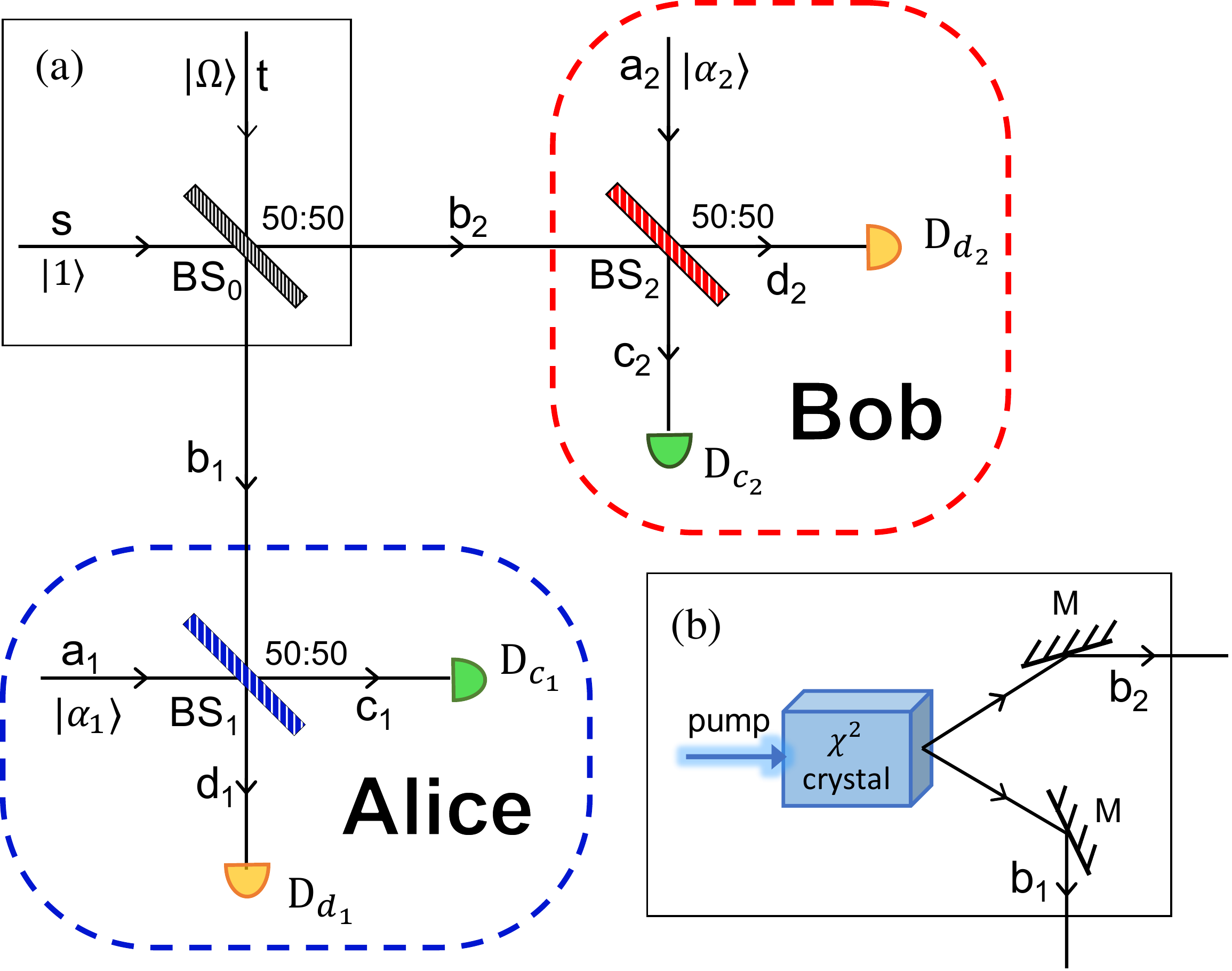}
	\caption{\label{mainSetup}
	Experimental configuration (a) proposed  by Tan, Walls and Collett in \cite{TWC91}.  A single photon impinges on a 50-50 beamsplitter via input $s$, along with the vacuum in the input $t$. As a result we get state $\ket{\psi}_{b_1, b_2}$, which propagates to the laboratories of Alice and Bob, who perform homodyne  measurements involving weak coherent local oscillator fields (their amplitudes satisfy $|\alpha_1|=|\alpha_2|$), and photon number resolving detectors D. In experimental configuration (b), modes $b_1$ and $b_2$ are fed with radiation in two-mode squeezed state (Grangier, Potasek, Yurke \cite{GRANGIER-P-Y}). Configuration ``on/off" (c) is the same as (b), with  local oscillator switched off for the local "off" setting, and beamsplitters of optimized transmittivity (e.g. 0.8 for (b)). } 
\end{figure}

{\em Experiment (a) \cite{TWC91}.} TWC considered the state obtained by casting a single photon on a 50-50 beamsplitter,
$
	\label{PSI}
	\ket{\psi}_{b_1, b_2} = \frac{1}{\sqrt{2}} ( \ket{01}_{b_1, b_2}+ i \ket{10}_{b_1, b_2} ),
$
where e.g. $|10\rangle_{b_1,b_2},$ indicates one photon excitation in 
exit mode $b_1$ and  vacuum in
exit mode $b_2$, see Fig. (\ref{mainSetup}). They suggested that its Bell non-classicality can be revealed in weak-field homodyne measurements. No photon number resolution was assumed at the time.
The form of $\ket{\psi}_{b_1, b_2}$ appears to be similar to a two-qubit Bell state. However, $\ket{\psi}_{b_1, b_2}$ is intrinsically different in terms of the number of particles involved and can be also thought of as a plain superposition of the photon in either of the beams. 

In the experimental proposal of  TWC, 
Fig. (\ref{mainSetup}) (a), the state $\ket{\psi}_{b_1,b_2}$ is distributed between Alice (controlling the mode $b_1$) and Bob ($b_2$).  They both perform weak-field homodyne measurements on their parts. Modes $b_j$, where $j = 1,2$, and the auxiliary coherent fields in state $\ket{\alpha_j}_{a_j}=\ket{\alpha e^{i \theta_j}}_{a_j}$ are fed into input ports of  50-50 beamsplitters BS$_j$,  and end up in detectors $D_{c_j}$ and $D_{d_j}$. 

The analysis of TWC was using the run-of-the-mill of the time  approach to photodetection, according to which the probability of a detector to fire is proportional to the intensity of the impinging light, e.g. \cite{LOUDON}. They modelled the intensity observable by the photon number operator. Their discussion assumes that the joint probability of having a coincidence of firings for  detectors 
$D_{x_1}$ and $D_{y_2}$  is $P_f(x_1,y_2|\theta_1, \theta_2)\approx \av{I_{x_1}(\theta_1)I_{y_2}(\theta_2)}$.
Here $I_{x_j}(\theta_j)$ is the intensity at output $x=c,d$,
and the averaging is done, depending on the context, over local hidden variables,  or within quantum formalism.
Without assuming photon number resolution, the probability of a single firing reads:  
 $P_f(x_j|\theta)\approx \av{I_{x_j}(\theta)}.$

To show a violation of local realism, TWC used the   correlation functions
   \begin{equation} \label{CORR-FUNCTION}
   \!  E(\theta_1, \theta_2) = \frac{\langle (I_{c_1}(\theta_1)-I_{d_1}(\theta_1))(I_{c_2}(\theta_2)-I_{d_2}(\theta_2))\rangle}{\langle(I_{c_1}(\theta_1)+I_{d_1}(\theta_1))(I_{c_2}(\theta_2)+I_{d_2}(\theta_2))\rangle}.
   \end{equation}
    and following \cite{Reid86} a Bell-like inequality for these of the well known CHSH form \cite{CHSH}. 
The local settings were defined by the local phases $\theta_j$ and $\theta'_j$. For (constant) amplitudes of the local oscillators satisfying $\alpha^2<\sqrt2 -1$, they showed a violation of the CHSH-like  inequality, and concluded that the single-photon state $\ket{\psi}_{b_1,b_2}$
is ``nonlocal''. 
But, the inequality, derived in\,\cite{Reid86}, rests on an {\em additional} assumption that in LHV models
the total intensity for each observer $j$ does {\em not} depend on $\theta_j$:
$\label{ASSUMPTION}
I_j(\lambda)=I_{c_j}(\theta_j, \lambda)+I_{d_j}(\theta_j, \lambda), $
where $\lambda$ symbolizes the hidden variables. 
As observed in \cite{Santos92}  and \cite{Zukowski16} such a condition is justified in classical optics, but severely constrains possible LHV models. 

Santos suggested that the correlations of firings
$P_f(x_1,y_2|\theta_1, \theta_2)$ in the TWC scheme {\em could   be} explainable 
with local hidden variables  \cite{Santos92} and therefore cannot be used to convincingly demonstrate  ``nonlocality'' of a single photon. However, his LHV model reproduced only the correlations of firings, and not the full quantum predictions, failing completely 
to recover quantum predictions for $P(x_j|\theta)\approx \av{I_{x_j}(\theta)}$. Thus, this was not a hidden variable model of the quantum predictions, 
but rather a hint that there is something wrong in the TWC analysis. Santos 
suggested that the additional assumption   is violated rather than local realism.
One has to add that  Santos 
did {\em not} consider photon number resolving detection.

Other works challenged the single-photon nature of the effect \cite{Greenberger95, Peres95}, or suggest modifications of the experiment which would allow provable violations of local realism \cite{Hardy94, Banaszek99}, while keeping the experiment all-optical. Thus far, no definite answer was given to the problem whether the TWC interference effect, which seemingly violates local realism,
admits a precise local realistic model, or not. 
Papers describing the experimental realizations of variants of this scheme  \cite{KUZMICH, Hessmo04, Babichev04} claim  violations of a Bell inequality. However, these claims were presented with caution, e.g in \cite{Babichev04}, where it is stated that the results are no better than those for conventional Bell tests with the efficiency loophole.

Below we show a LHV model which reproduces precisely quantum predictions for the TWC setup (a), even in the case of photon number resolving detection. This obviously covers the course-grained description in terms of the probabilities of firings of detectors whose response is proportional to the  number of impinging photons.
Its applicability is limited by the strength of the local oscillators, but covers the range reported in \cite{TWC91} as revealing the ``nonlocality of the single-photon''.
The result closes the case and precludes any attempt to implement device-independent protocols using TWC correlations.

{\it Explicit LHV model of TWC correlations.} Quantum predictions for the TWC setup are fully characterized by the probabilities $p(\mathbf{n})$ of events consisting of registering a specific numbers of photons in the  output modes: $\mathbf{n}=(k_{c_1},l_{d_1},r_{c_2},s_{d_2})\in\mathbb{N}^4$  (for readability, we omit the indices indicating the modes in further parts of this report). They read (see Appendix A for the derivation): 
\begin{equation}
p(\mathbf{n})=A(\alpha,\mathbf{n})\Big[ (k\!-\!l)^2 +(r\!-\!s)^2+ 2(k\!-\!l)(r\!-\!s)\sin(\theta_{12}) \Big],
\label{eq:probability}
\end{equation} 
where  $A(\alpha,\mathbf{n})=\frac{e^{-2\alpha^2} \left(\frac{\alpha^2}{2}\right)^{k+l+r+s}}{ 2\alpha^2 k!~ l!~ r! ~ s!} $ and $\theta_{12}=\theta_1-\theta_2$.

Note that, whenever both detectors of Alice {\it or} Bob register the same number of photons, the probability does not depend on $\theta_{12}$. Let us denote the set of these events as $\mathcal{N}:=\{\mathbf{n} : k=l\,\, \text{or}\,\, r=s\}.$ 
We cover them by a family of trivial LHV submodels assigning fixed outcomes to Alice and Bob, see further.

Next, notice that all the probabilities that do depend on $\theta_{12}$ are of the form
 \begin{equation}
 \label{prob_int}
    p(\mathbf{n})=B(\alpha,\mathbf{n})\,( 1+\mathcal{V}(\mathbf{n})\sin(\theta_{12}) ),
\end{equation}
where $B(\alpha,\mathbf{n})=A(\alpha,\mathbf{n})\Big[ (k\!-\!l)^2 +(r\!-\!s)^2\Big]$ and
  $ \mathcal{V}(\mathbf{n})~=~\frac{2(k\!-\!l)(r\!-\!s)}{ (k\!-\!l)^2 +(r\!-\!s)^2}.$
To reproduce them, we adapt a model by Larsson \cite{LARSSON99}, see also \cite{Franson99}, which reproduces fully quantum predictions for a two-qubit singlet state, provided that the detection inefficiency is lower than $2/\pi$

Our model $\mathcal{M}$ is a convex combination of submodels $\mathcal{M}_\mathbf{n}$, each chosen with
probability $P(\mathcal{M}_{\mathbf{n}})$. The submodels belong to two infinite families: the trivial $\{\mathcal{M}_\mathbf{n}\}_{\mathbf{n}\in\mathcal{N}}$ and Larsson-like one $\{\mathcal{M}_\mathbf{n}\}_{\mathbf{n}\in\mathcal{\tilde{N}}}$, where $\mathcal{\tilde{N}}:=\{\mathbf{n} : k>l\,\, \text{and}\,\, r>s\}$. 
We focus on the latter first.

We group the probabilities that depend on the local settings $\theta_j$ with the ones that correspond to the events in which (perfect) detectors of either Alice or Bob do not register any photons, 
and denote the set of such events by $\mathcal{O}:=\{\mathbf{n}\in\mathcal N : k=l=0\,\, \text{or} \,\,r=s=0\}$.

Each Larsson-like submodel $\{\mathcal{M}_{(k,l,r,s)}\}$ is going to predict eight events resulting from applying (or not) the swaps $k\leftrightarrow l$ and $r\leftrightarrow s$ to events $(0,0,r,s)$, $(k,l,0,0)$ and $(k,l,r,s)$. Notice that only one of the above 
matches the \emph{index} $(k,l,r,s)\in\tilde{\mathcal N}$ of the model.
To construct it, we take a  uniformly distributed continuous hidden variable $\lambda\in[0,2\pi]$ and a coin toss one $x\in\{0,1\}$.

Specifically, for $x=0$  Alice can register the event $(c,d)\in\{(k,l),(l,k)\}$ with probability
\begin{eqnarray}
\label{janoke}
  &\!\! \! \!\!\!\!P_{\mathbf{n}}^A(c,d|\theta_1,\lambda,0)=R_{\mathbf{n}}(c,d|\theta_1, \lambda)= \frac{1-\mathcal{V}(\mathbf{n})}{\pi}\nonumber \\ &\!\!\!\!\!\!\!+\mathcal{V}(\mathbf{n})|\sin{(\theta_1-\lambda)}| H_{}\left((c-d)\sin(\theta_1-\lambda)\right),
  \end{eqnarray}
  where $H$ is the Heaviside function, and for $(0,0)$ event we put
\begin{eqnarray}
\label{ProbA00}
  P_{\mathbf{n}}^A(c=0,d=0|\theta_1,\lambda,0) = R_{\mathbf{n}}(0,0|\theta_1,  \lambda)
  \nonumber \\
  =1-\sum\limits_{(e,f)\in\{(k,l),(l,k)\}}R_{\mathbf{n}}(e,f|\theta_1, \lambda).
\end{eqnarray}
Bob detects $(c',d')\in\{(r,s),(s,r)\}$ with probabilities 
\begin{eqnarray}
\label{janoke3}
P_{\mathbf{n}}^B(c',d'|\theta_2,\lambda,0)=Q_{\mathbf{n}}(c',d'|\theta_2, \lambda)\nonumber \\ =H\left((c'-d')\cos(\theta_2-\lambda)\right). 
\end{eqnarray}
For $x=1$ we put 
$P_{\mathbf{n}}^A(c,d|\theta_1, \lambda, 1)= Q_{\mathbf{n}}(c,d|\theta_1, \lambda)$, 
and $P_{\mathbf{n}}^B(c',d'|\theta_2, \lambda, 1)= R_{\mathbf{n}}(c',d'|\theta_2, \lambda).$ This symmetrizes the model.

The predictions for joint probabilities specified by each submodel are given by:
$\!\!P^{AB}_{\mathbf{n}}(c,d,c',d'|\theta_1, \theta_2) \!
\!\! =\frac{1}{{4\pi}}\sum\limits_{x=0}^1 \int_{0}^{2\pi}\!\! \!\!\!
d\lambda\,  P^A_{\mathbf{n}}(c,d|\theta_1, \lambda,x)\,P^B_{\mathbf{n}}(c',d'|\theta_2, \lambda,x). $
The explicit probability that the submodel $\mathcal{M_{\mathbf{n}}}$ predicts the event $\mathbf{n}=(k,l,r,s)$ in the simplest case of  $\pi/2> \theta_1  > \theta_2>0$, $k>l$ and $r>s$, reads

$ P^{AB}_{\mathbf{n}}(k,l,r,s|\theta_1, \theta_2)=\frac{1+\mathcal{V}(\mathbf{n})\sin{\theta_{12}}}{2\pi}.$
All other predictions of the submodel can be obtained similarly.
For  events 
$(k,l,r,s),(l,k,r,s),(k,l,s,r)$ and $(l,k,s,r)$ we get a concise formula:
\begin{equation}
\label{interference}
P^{AB}_{\mathbf{n}}(c,d,c',d'|\theta_1, \theta_2) =\frac{1\!+\mathcal{V}(\mathbf{n})\,\text{sgn}\left((c\!-\!d)(c'\!-\!d')\right)\sin(\theta_{12})}{2 \pi}.
\end{equation}
In the case of the $\mathcal O$-events $(0,0,r,s),(0,0,s,r),(k,l,0,0)$ and $(l,k,0,0)$, the probability is flat and reads $\frac{1}{4}- \frac{1}{2\pi},$
which follows directly from the normalisation condition in Eq.  \ref{ProbA00}).
Comparing \eqref{interference} with the corresponding quantum probabilities \eqref{prob_int}, we see that each Larsson-like submodel $\mathcal{M}_{\mathbf{n}}$ must appear in the  full model $\mathcal{M}$ with probability $P(\mathcal{M}_{\mathbf{n}})=2\pi B(\alpha,\mathbf{n})$.

In Appendix B we show that  formulas for $ P(\mathcal{M}_{\mathbf{n}})$ lead to a properly normalized probability distribution, with a proviso described below.

The  model reproduces all probabilities which reveal interference.
However a {\em condicio sine qua non}  for consistency of the full model is to properly describe  also  events 
$\mathcal{O}$. The construction, due to the Larsson-like submodels, ascribes probability $(\frac{\pi}{2}-1) B(\alpha,(k,l,c',d'))$ to the event $(k,l,0,0)$ and $(l,k,0,0)$. This is so, because  for each of the  submodels  $\mathcal{M}_{(k,l,c',d')}$,
a fraction $\frac{1}{4}- \frac{1}{2\pi}$ of it covers these events from $\mathcal O$, and the sub-model as a whole appears with probability $2\pi B(\alpha,\mathbf{n})$.

The sum of all such contributions cannot be greater than the quantum probability $p(k,l,0,0)$. It can be lower since the difference can be described by trivial models. This gives the following consistency conditions:
\begin{eqnarray}
\label{condition}
&&\Delta_{(k,l,0,0)}=
p(k,l,0,0)\nonumber \\
&&-\left(\frac{\pi}{2}-1\right)\sum\limits_{c'> d'} B(\alpha,(k,l,c',d'))\geq 0,
\end{eqnarray}
which must hold for any $k\neq l$.
Due to the symmetrization an analogous constraint holds for events of $(0,0,r,s)$.

In Appendix C we show that the condition in Eq. (\ref{condition}) is satisfied for 
{\em any} $(k,l)$ and $(r,s)$, whenever $\alpha^2<0.87$. 
The model can be completed using a family of trivial submodels $\mathcal{M}_{\mathbf{n}}$ for events $\mathbf{n}\in\mathcal{N}$. They predict fixed outcomes for Alice and Bob, $P^A_{\mathbf{n}}(k,l)=P^B_{\mathbf{n}}(r,s)=1$, which lead to $P^{AB}_\mathbf{n}(k,l,r,s)=1$. 
Obviously, for events $\mathbf{n}\in\mathcal{N}\setminus{\mathcal{O}}$, we choose each corresponding trivial model $\mathcal{M_{\mathbf{n}}}$ with probability $p(\mathbf{n})$. Finally, for events $\mathbf{n}\in\mathcal{O}$ we might need to
compensate the potential difference $\Delta_{(k,l,0,0)} > 0$ between the quantum predictions for the $\mathcal{O}$-events and the predictions specified by the Larsson-like models. To do that, we use an additional  trivial submodel for event ${(k,l,0,0)}$, which appears in the full model  with probability $P(\mathcal{M}_{(k,l,0,0)})=\Delta_{(k,l,0,0)}$.
The case of $\Delta_{(0,0,r,s)}>0$ is treated the same way. 
 
One can easily build a better version of the model which would hold for slightly higher values of $\alpha$. However, we were not able to find a model which has an unconstrained validity, and one can 
conjecture that the Larsson-like approach cannot lead to such.  Still, our model fully covers the range of $\alpha$ for which TWC
predicted a violation of local realism. Thus, this claim is fully revoked,  and this is done for  finer, photon-number resolving, measurements than in the original proposal.

Importantly, the model covers the range of local oscillator amplitudes $\alpha$ which can be thought of as giving weak-field homodyne measurements (which following \cite{WEAKHOMO-OPTICAL-COHERENCE} would  mean here photon numbers, $\alpha^2$, in the local oscillators close to 1). Even more importantly the model works for the intensities of the local oscillators used in \cite{WEAKHOMO-OPTICAL-COHERENCE}, Fig. 3 there, given by $\alpha\approx 0.55$.  

{\it Experiment (b).}
The setup proposed in \cite{GRANGIER-P-Y}, and realised in \cite{WEAKHOMO-OPTICAL-COHERENCE} in a weak-field homodyne photon-number resolving version, is a modification of the one proposed in  Fig. \ref{mainSetup}. In the setup the source  is a parametric down conversion process, in which a non-linear crystal transforms part of the pumping light into a pair of photons, fed into two output modes, $b_1, b_2$. The photons are sent to two  measurements stations which perform weak (in the version of \cite{WEAKHOMO-OPTICAL-COHERENCE} photon-number resolving) homodyne measurements.
The state of beams $b_1$ and $b_2$ is a two mode squeezed vacuum, 
$
   \ket{\sigma_v}= \sqrt{1-\gamma^2}\sum_{k=0}^{+\infty}  \gamma^k \ket{k, k}, 
$
where $\gamma$ is for simplicity assumed to be real. As before, Alice and Bob mix $ \ket{\sigma_v}$  with 
two weak coherent states in modes $a_1$ and $a_2$, namely $\ket{\alpha_j}_{a_j} = \ket{\alpha e^{i \theta_j}}_{a_j} $  using local beamsplitters.

Below we show an explicit model for a subset of correlations appearing in the GPY setup with photon-number resolution, which is based on the model for the TWC correlations, and covers events  with maximally one photon  detected on one side, and maximally three on the other one. They depend only on $\theta_1-\theta_2$. We call them \emph{class 1} events.  We were not able to find an explicit Larsson-model for events with two photons detected on both sides, as they depend also on $2\theta_{12}$. Still we show,  that there is no evidence that such a model does not exist.
This is done by showing that  CGLMP inequalities \cite{CGLMP02} cannot be violated by this subset of events.
Thus, for all events with altogether up to four detected photons there is no evidence for violation of local realism. As these events definitely are most frequent, this constitutes a  strong argumentation towards showing that the GPY configuration does not constitute a proper Bell test. 

\emph{Class 1} events follow a pattern: one party registers a local event (k,l), $k\neq l$, while the other detects a single photon. Their probabilities are of the form $ p(\mathbf{n})=A(\mathbf{n})\,\left(\alpha^4+c_1(\mathbf{n})\,\gamma^2 \pm c_2(\mathbf{n})\,\alpha^2\gamma  \cos \theta_{12}\right)$ (summarized in Appendix E).
Larrson like models exist in this case.
In each of them, one party predicts local events (k,l) and (l,k) with probability $R(c_1(\mathbf{n}),c_2(\mathbf{n}),\pm)$
    and (0,0) w.p. $1-R(c_1(\mathbf{n}),c_2(\mathbf{n}),+)-R(c_1(\mathbf{n}),c_2(\mathbf{n}),-)$, where
\begin{eqnarray}
    R(c_1,c_2,\pm)=\frac{1-c_2\alpha^2 \gamma}{\pi(\alpha^4+c_1\gamma^2)}\nonumber\\ +\frac{c_2\alpha^2 \gamma}{(\alpha^4+c_1\gamma^2)}|\cos{(\theta_1-\lambda)}|\, H_{}\left(\pm\cos(\theta_1-\lambda)\right).
\end{eqnarray}         
    The other party predicts (1,0) and (0,1) with probability $H_{}\left(\pm\cos(\theta_1+\lambda)\right)$.

Submodels are chosen with probability $2\pi A(\mathbf{n})\,c_1(\mathbf{n})$. The events whose probabilities do not depend on local settings are covered by trivial models, with the exception of single-photon events. This is because the submodels corresponding to \emph{class 1} events contribute to their probabilities. Just as in the TWC case, we need to check if the sum of these contributions does not exceed the quantum probabilities. This provides a condition for the validity of the model. In the appendix D we show that in the case of 
$\gamma\leq \alpha^2$ it restricts the model to $\alpha^2<0.58$, but this does not characterize its full range, which is broader.

The remaining  events with two photons registered on both sides, which we shall denote as $(2\&2)$, come from the following term of the expansion of the overall state (PDC modes plus local oscillators),
$ |\xi_{2\&2}\rangle \!=\!
Z \left(\frac{\alpha_1^2\alpha_2^2}{4}  a_1^{\dagger2} a_2^{\dagger2} \!
+ \!\gamma\alpha_1\alpha_2 a_1^\dagger a_2^\dagger b_1^\dagger b_2^\dagger+ \frac{\gamma^2}{2}b_1^{\dagger2} b_2^{\dagger2}\right) |{\Omega}\rangle, $
where $Z$ is the overall normalization constant of the entire state.
The overall  probabilistic weight of this term is thus $p(2\&2)=\frac{Z^2}{4}(\alpha^8+4\gamma^4 +4\gamma^2\alpha^4)$.
The events in the case of which two photons are registered on one side and on the other side no photons, denoted by $(2\&0)$, come from the component  $|\xi_{2\&0}\rangle= Z\Big(\frac{1}{2!} \alpha_1^2 a_1^{\dagger2} + \frac{1}{2!} \alpha_2^2 a_2^{\dagger2}\Big) |{\Omega}\rangle. $ Its overall probabilistic  weight is $p(2\&0)=Z^2\alpha^4$.

Note that $|\xi_{2\&2}\rangle$, after normalization,
is a proper 3 dimensional state, and thus predictions for it can be put into a CGLMP inequality for $d=4$. 
We choose $d=4$ because we want to analyze it together with $|\xi_{2\&0}\rangle$. We ascribe the following numeric values to the results: for Alice and Bob, counts $00$, $02$, $20$, $11$, are assigned values $a,b=0,1,2,3$. 
The optimal way of ascribing these values is given below. 

The overall probability of  events  $(2\&2)$ and$(0\&2)$ is a convex combination of the two cases. For the considered subset, S, of events 
$p(a,b|S)= \lambda p(a,b|2\& 2) + (1-\lambda) p(a,b|0 \& 2),$
with
$\lambda=\frac {P(2\&2)}{P(2\&2)+P(0\&2)}.$

The CGLMP expression is linear. Thus its value, $W[p(\cdot|S)]$, for the  convex combination reads 
$
\lambda W[ p(\cdot|2\&2)]
+(1-\lambda)W[ p(\cdot|0 \& 2)].
$
The maximal possible {\em algebraic} value of the CGLMP expression is $4$.
Thus $W[ p(\cdot|2\&2)]\leq4$.
By considering the explicit form of the CGLMP inequality we see that the maximal possible value for $W[ p(\cdot|0 \& 2)]$ is
$2\left[p(a=b|0\&2)-\frac{1}{3}p(a\neq b|0\&2)\right]=\frac{8}{3}p(a=b|0\&2)-\frac{2}{3}.$
This is because the probabilities $p(a,b|0 \& 2)$ are independent of the settings.
We must seek such a function relating photon counts $00$, $02$, $20$, $11$ with these numbers $a,b$, such that the value of $W[ p(\cdot|0 \& 2)]$ is highest.
This is so when e.g $00$ result on Alice's side  is ascribed $0$ and $11$ on Bob's side is also $0$, and $11$ on Alice's is given 1 and $00$ on Bob's also 1. Then $W[ p(\cdot|0 \& 2)]=\frac{2}{3}$, because $p(a=b)=\frac{1}{2}$. Thus the value of the CGLMP expression cannot be higher than
$ W[ p(\cdot|S)]= 4\lambda +(1-\frac{2}{3}\lambda)= \frac{7}{3}\lambda+\frac{2}{3}.$
As the local realistic bound is 2, {\em a} necessary condition for having a local realistic description for the considered probabilities is 
$W[ p(\cdot|S)]=4\lambda +\frac{2}{3}(1-\lambda)= \frac{10}{3}\lambda+\frac{2}{3}, $ 
which does not breach $2$ for $\lambda\leq 0.4$, and thus for $\alpha^2=\gamma$ smaller than  approximately $0.54$.  Since $\lambda(\gamma^2)$ increases together  with $\gamma^2$, this holds also for all $\gamma< \alpha^2<0.54$. 
This range is consistent with the range of the Larsson like models, especially as for the state $|\xi_{(2\&2)}\rangle$
one definitely must have $W[ p(\cdot|2\&2)]\leq4$, since the Tsirelson bound for $d=4$ CGLMP inequality is much less than 4. In ref. \cite{LASKOWSKI} it is estimated to be around $3.1$.

Thus, there is \emph{no evidence} in terms of a violation of CGLMP Bell inequality that \emph{there is no local realistic model} for experiments of the type \cite{WEAKHOMO-OPTICAL-COHERENCE}, and their precursors like \cite{KUZMICH}, at least for the situation in which up to two photon counts per party matter. We conjecture that this can be extended to more-photon events. We have chosen the CGLMP inequality as it appears in \cite{WEAKHOMO-OPTICAL-COHERENCE} in an argument on Bell nonclassicality of the correlations. Of course, our claim is not that the model exists, but that there is currently no evidence that it does not exist, and that a partial explicit model exist. Still, this is not the end of the story for photon-number resolving homodyne measurements in the context of violation of Bell inequalities, as shown below.  

{ \it On/off unbalanced homodyne measurements as a solution.}
We use here CH inequality \cite{CH74}: 
 \begin{eqnarray}         \label{CHin}
   -1~\leq~P(A,B)+P(A,B')+P(A',B) \nonumber \\-P(A',B') 
        -P(A)-P(B)~=~CH~\leq~0. 
   \end{eqnarray}
For the on/off arrangement by event $A'$ we denote a single photon detected in mode $d_1$ and no-photon in mode $c_1$ in the case we have a beamsplitter (now of an optimized transmittivity $T$) and the local oscillator field on (`on' setting), and by $A$ we denote a single photon count at either $d_1$ or $c_1$ when the local oscillator is off (`off' setting). Events $B$ and $B'$ play the same role for Bob.

\emph{Experiment (a)}. 
For $\alpha_1=i\alpha_2$ we have 
$ P(A',B') = 2\alpha^2 e^{-2\alpha^2} T(1 - T)$. The other probabilities are $P(A',B) = P(A,B') = \frac 12 \alpha^2  e^{-\alpha^2}(1 - T),$ 
and for the (trivial) off/off ones are $P(A,B) = 0,$ and $P(A) = P(B) = \frac 12.$ 
The condition for $CH < -1$ reads $\frac 12 e^{\alpha^2} < T < 1$.
An optimization results in $CH_{\min} \approx -1.010$, for $\alpha^2 \approx 0.196, \, T \approx 0.804$. The violation is quite robust as the probabilities for the non-trivial case read only $P(A',B) = 0.0157,$  $P(A',B') = 0.0417$.

 \emph{Experiment (b).}
 We have $ P(A',B')=e^{-2\alpha^2} (1-\gamma^2) (T \gamma - \alpha^2 (1 - T))^2$, whereas  $ P(A',B)=P(A,B')=e^{-\alpha^2}\gamma^2 (1-\gamma^2) T $, finally   $P(A,B) = P(A) = P(B)  = \gamma^2(1-\gamma^2) $. The necessary condition for $CH>0 $ reads   $ T~>~\frac12.$ Optimization of the CH expression yields $ CH_{\max} \approx 0.0027 $ 
 for $\alpha^2 \approx 0.200$, $\gamma \approx 0.175$ and $T \approx 0.799.$
The violation is quite robust as $P(A,B)\approx 0.0299$, \,$P(A,B')\approx0.0196$ and $ P(A',B')\approx 0.0065.$ 

{\it Closing remarks.} 
We see that the TWC configuration and most probably the GPY scheme are not proper Bell experiments.
Still, if one resorts to measurement settings involving on/off local oscillators and optimized non $50-50$  beamsplitters a violation of local realism is detectable. This works for both the TWC configuration (a) 
and the GPY one (b).

The on/off scheme obviously represents the most extreme version of complementarity between measuring wave aspects of the state of the modes vs the particle ones. Thus, it seems that quantum optical Bell tests with homodyne measurements must involve at least one operational situation in which photon counting replaces weak-field homodyne measurements. This does work also when the local oscillator is almost off, see for the TWC case \cite{2ndPaper}. 
On/off situation led to a violation of local realism in Refs. \cite{Hardy94} and \cite{Banaszek99}, albeit in a slightly different situations: modification of the signal state in beam $s$ in \cite{Hardy94}, and a displacement procedure in \cite{Banaszek99}. Our results suggest that this is not a peculiarity, but seems to be a rule for homodyne photon number resolving  measurements. 

{\em Acknowledgements.} Work supported by  Foundation for Polish Science (FNP), IRAP project ICTQT, contract no. 2018/MAB/5, co-financed by EU  Smart Growth Operational Programme. MK is supported by  FNP  START scholarship. AM is supported by (Polish) National Science Center (NCN): MINIATURA  DEC-2020/04/X/ST2/01794.

\bibliography{SinglePhoton}
 
\onecolumngrid

\appendix 

\section*{Appendix A -- Experiment (a): Quantum photodetection probabilities}
In this section, we are going to calculate the probability of detecting the event $n=(k_{c_1},l_{d_1},r_{c_2},s_{d_2})$, consisting in registering  specific numbers of photons in the output modes of the Tan-Walls-Collett setup.

The initial state in the TWC scheme, obtained by transforming a single photon with a balanced beamsplitter and adding two coherent states of the local oscillators
\begin{equation}
\label{a1}
|\Psi\rangle=|\alpha e^{i \theta_1}\rangle_{a_1}  \frac{1}{\sqrt{2}} (|01\rangle+i\,|10\rangle)_{b_1b_2}|\alpha e^{i \theta_2}\rangle_{a_2}.
\end{equation}

We show how the state \eqref{a1} transforms on balanced beamsplitters $U_{BSj}, ~j = 1,2$ which link the output and input modes via
\begin{equation}
\label{a2}
   \hat c_j=\frac{1}{\sqrt2}(\hat a_j+i \, \hat b_j) \,\, \text{and} \, \,  \hat d_j=\frac{1}{\sqrt2}(i\,\hat a_j+ \hat b_j).
\end{equation}
Applying \eqref{a2} to the state \eqref{a1} we get
\begin{eqnarray}
\ket{\Psi} &=& e^{-\alpha^2 } \sum_{j=0}^{\infty}  \frac{(\alpha e^{i \theta_1})^j}{j!} (\hat a_1^\dagger)^j \frac{1}{\sqrt{2}} ( i\hat b_1^\dagger  + \hat b_2^{\dagger})   \sum_{k=0}^{\infty}  \frac{(\alpha e^{i \theta_2})^k}{k!} (\hat a_2^\dagger)^k \nonumber\\
&=& e^{-\alpha^2 }\sum_{j,k=0}^{\infty} {2}^{-\frac{j+k}{2}}\frac{(\alpha e^{i \theta_1})^j}{{j!}}\frac{(\alpha e^{i \theta_2})^k}{{k!}}\Big( \hat c_1^{\dagger} +i\hat d^{\dagger}_1 \Big)^j\frac 12 \Big(-\hat c_1^{\dagger} + i\hat d_1^\dagger  +i \hat c_2^\dagger + \hat d_2^{\dagger}\Big)  \Big( \hat c_2^{\dagger} + i\hat d^{\dagger}_2 \Big)^k\ket{\Omega} \nonumber \\
&=&  e^{-\alpha^2 }\sum_{j,k=0}^{\infty} {2}^{-\frac{j+k}{2}}\frac{(\alpha e^{i \theta_1})^j}{{j!}}\frac{(\alpha e^{i \theta_2})^k}{{k!}}\sum_{p=0}^j\binom{j}{p}(\hat c^{\dagger}_1)^{j-p}(i\hat d_1^{\dagger})^{p}\frac 12 \Big(-\hat c_1^{\dagger} + i\hat d_1^\dagger  +i \hat c_2^\dagger + \hat d_2^{\dagger}\Big)  \sum_{q=0}^k\binom{k}{q}(\hat c^{\dagger}_2)^{k-q}(i\hat d_2^{\dagger})^{q} \ket{\Omega},\nonumber \\
&=& \sum_{j,k=0}^{\infty}  \sum_{p=0}^j \sum_{q=0}^k f(j,p,k,q) (\hat c^{\dagger}_1)^{j-p}(\hat d_1^{\dagger})^{p} \Big(-\hat c_1^{\dagger} + i\hat d_1^\dagger  +i \hat c_2^\dagger + \hat d_2^{\dagger}\Big)(\hat c^{\dagger}_2)^{k-q}(\hat d_2^{\dagger})^{q} \ket{\Omega}, \\
&=&  \sum_{j,k=0}^{\infty}\sum_{p}^{j}\sum_{q=0}^kf(j,p,k,q)\bigg[ - \sqrt{(j-p+1)!p!(k-q)!q!}\ket{j-p+1}_{c_1}\ket{p}_{d_1}\ket{k-q}_{c_2}\ket{q}_{d_2}  \nonumber \\
&& + i\sqrt{(j-p)!(p+1)!(k-q)!q!}\ket{j-p}_{c_1}\ket{p+1}_{d_1}\ket{k-q}_{c_2}\ket{q}_{d_2}\nonumber \\
&& + i\sqrt{(j-p)!p!(k-q+1)!q!}\ket{j-p}_{c_1}\ket{p}_{d_1}\ket{k-q+1}_{c_2}\ket{q}_{d_2} \nonumber \\
&& +  \sqrt{(j-p)!p!(k-q)!(q+1)!}\ket{j-p}_{c_1}\ket{p}_{d_1}\ket{k-q}_{c_2}\ket{q+1}_{d_2}  
\bigg] 
\end{eqnarray}
where 
\begin{eqnarray}
f(j,p,k,q) &=& e^{-\alpha^2 } {2}^{-\frac{j+k}{2}-1} \frac{(\alpha e^{i \theta_1})^j}{{j!}}\frac{(\alpha e^{i \theta_2})^k}{{k!}} \binom{j}{p} \binom{k}{q}  (i)^{p+q}, ~~~~ \forall p \leq j, q \leq k.
\label{gFUNCTION}
\end{eqnarray} 
Now, 
\begin{eqnarray} 
&& \text{Pr}(k, l; r, s) = |\bra{k,l,r,s}\ket{\Psi}|^2 \nonumber \\
&=& \bigg|
- f(k+l-1,l,r+s,s) + i f(k+l-1,l - 1,r+s,s)  \nonumber \\
&& \hspace{4cm} +i f(k+l,l,r+s-1,s)
+f(k+l,l,r+s-1,s-1)\bigg|^2  k!~ l!~ r! ~ s! \nonumber \\
&=& \frac{e^{-2\alpha^2}}{ k!~ l!~ r! ~ s!} \Big(\frac{\alpha^2}{2}\Big)^{k+l+r+s} \frac{1}{2\alpha^2}  \Big[ (k-l)^2 + (r-s)^2 + 2(k-l)(r-s) \sin(\theta_1 - \theta_2) \Big], ~~
\label{eqapp:probability}
\end{eqnarray} 

\section*{Appendix B -- Experiment (a): Explicit calculation of the sum of probabilities of all submodels $\mathcal{M_\mathbf{n}}$}

In this section we prove that the probabilities $P(\mathcal{M_\mathbf{n}})$ of choosing specific submodels are properly normalized. We have
\begin{eqnarray}
\label{b1}
    \sum\limits_{\mathbf{n}\in\mathcal{N}\cap\mathcal{\tilde{N}}}P(\mathcal{M_\mathbf{n}})=\sum\limits_{\mathbf{n}\in\mathcal{N}\setminus\mathcal{O}}B(\alpha,\mathcal{M_\mathbf{n}})+\sum\limits_{\mathbf{n}\in\mathcal{\tilde{N}}}2\pi B(\alpha,\mathcal{M_\mathbf{n}})+\sum\limits_{\mathbf{n}\in\mathcal{O}}\Delta_\mathbf{n},
\end{eqnarray}
where

\begin{eqnarray}
\label{b2}
  \sum\limits_{\mathbf{n}\in\mathcal{O}}\Delta_\mathbf{n}=
\sum\limits_{k\neq l}  \left(p(\mathbf{(k,l,0,0)})-\left(\frac{\pi}{2}-1\right)\sum\limits_{c'> d'} B(\alpha,(k,l,c',d'))\right)+
\sum\limits_{r\neq s}  \left(p(\mathbf{(0,0,r,s)})-\left(\frac{\pi}{2}-1\right)\sum\limits_{c> d} B(\alpha,(c,d,r,s))\right)\nonumber\\
=\sum\limits_{\mathbf{n}\in\mathcal{O}}B(\alpha,\mathcal{M_\mathbf{n}})-4\left(\frac{\pi}{2}-1\right)\sum\limits_{\mathbf{n}\in\mathcal{\tilde{N}}} B(\alpha,\mathcal{M_\mathbf{n}})=\sum\limits_{\mathbf{n}\in\mathcal{O}}B(\alpha,\mathcal{M_\mathbf{n}})-\left(2\pi-4\right)\sum\limits_{\mathbf{n}\in\mathcal{\tilde{N}}} B(\alpha,\mathcal{M_\mathbf{n}}).
\end{eqnarray}
Moreover, notice that
\begin{equation}
\label{b3}
    \sum\limits_{\mathbf{n}\in\mathcal{\tilde{N}}} B(\alpha,\mathcal{M_\mathbf{n}})=\frac{1}{4} \sum\limits_{\substack{\mathbf{n}\in{\mathbb{N}^4}\\c\neq d,\, c'\neq d'}} B(\alpha,\mathcal{M_\mathbf{n}}).
\end{equation}
Plugging Eqs.(\ref{b2}) and (\ref{b3} )into Eq. (\ref{b1}) we get
\begin{eqnarray}
    &\sum\limits_{\mathbf{n}\in\mathcal{N}\cap\mathcal{\tilde{N}}}P(\mathcal{M_\mathbf{n}})=\sum\limits_{\mathbf{n}\in\mathcal{N}\setminus\mathcal{O}}B(\alpha,\mathcal{M_\mathbf{n}})+2\pi\sum\limits_{\mathbf{n}\in\mathcal{\tilde{N}}} B(\alpha,\mathcal{M_\mathbf{n}}) \nonumber \\
    &+\sum\limits_{\mathbf{n}\in\mathcal{O}}B(\alpha,\mathcal{M_\mathbf{n}})-\left(2\pi-4\right)\sum\limits_{\mathbf{n}\in\mathcal{\tilde{N}}} B(\alpha,\mathbf{n})=\sum_{\mathbf{n}\in{\mathbb{N}^4}}B(\alpha,\mathbf{n})=1.
\end{eqnarray}

\section*{Appendix C -- Experiment (a): Threshold intensity of local oscillators  for the validity of the LHV model for the TWC scheme}

In this section we prove that if $\alpha^2<0.87$, the probabilities of choosing a specific submodel $ P(\mathcal{M_\mathbf{n}})$ are non-negative. To do that, we only need to consider $\mathbf{n_0}\in\mathcal{O}$, for which $P(\mathcal{M}_\mathbf{n_0})=\Delta_\mathbf{n_0}$. Let us fix $\mathbf{n_0}=(k,l,0,0), \,k\neq l$, as the reasoning for $\mathbf{n_0}=(0,0,r,s)$ is fully analogous. We need to check the conditions in which 
\begin{eqnarray}
\label{c1}
\Delta_\mathbf{n_0}=B(\alpha,\mathbf{(k,l,0,0)})-\left(\frac{\pi}{2}-1\right)\sum\limits_{c'> d'} B(\alpha,(k,l,c',d')\geq 0.
\end{eqnarray}
We plug the definition of the function $B(\alpha,\mathbf{n})$ from the main text into \eqref{c1} and obtain, after some transformations,
\begin{eqnarray}
\Delta_\mathbf{n_0}=\frac{e^{-2 \alpha^2} 2^{-k-l-3} \left(\alpha^2\right)^{k+l-1} \left(-(\pi -2) e^{\alpha^2} \left(\alpha^2+(k-l)^2\right)+(\pi -2) I_0\left(\alpha^2\right) (k-l)^2+4 (k-l)^2\right)}{k!l!}.
\end{eqnarray}
It is easy to see that the condition $\Delta_\mathbf{n_0}\geq0$ is equivalent to
\begin{equation}
\label{c2}
    -(\pi -2) e^{\alpha^2} \left(\alpha^2+(k-l)^2\right)+(\pi -2) I_0\left(\alpha^2\right) (k-l)^2+4 (k-l)^2\geq0.
\end{equation}
As the Bessel function $I_0$ satisfies  $I_0\left(\alpha^2\right)\geq1$, the inequality \eqref{c2} can be approximated by a slightly stricter
\begin{equation}
    -(\pi -2) e^{\alpha^2} \left(\alpha^2+(k-l)^2\right)+(\pi -2)(k-l)^2+4 (k-l)^2=\left((\pi -2) \left(-e^{\alpha^2}\right)+\pi +2\right) (k-l)^2-(\pi -2) \alpha^2 e^{\alpha^2}\geq0.
\end{equation}
For $\alpha<1$, the coefficient $\left((\pi -2) \left(-e^{\alpha^2}\right)+\pi +2\right)$ standing in front of $(k-l)^2$ is positive. This means that the critical case we need to consider is $(k-l)^2=1$. Thus, we arrive at
\begin{equation}
\label{c3}
   \left((\pi -2) \left(-e^{\alpha^2}\right)+\pi +2\right) -(\pi -2) \alpha^2 e^{\alpha^2}\geq0.
\end{equation}
It can be shown that the inequality \ref{c3} is satisfied for \begin{equation}
    \alpha^2\leq W\left(\frac{2 e+e \pi }{\pi -2}\right)-1\approx 0.87,
\end{equation}
where $W$ denotes the Lambert $W$ function ($W(z)$ returns the principal solution for $w$ in $z=w e^w$ ).

\section*{Appendix D -- Experiment (b): Threshold intensity for the validity of the partial LHV model for the GPY scheme}
The presented model definitely reproduces all probabilities of events other than (0,0,0,1), (0,0,1,0), (0,1,0,0) and (1,0,0,0). We need to check if the probabilities of the single-photon events can also be recovered. To this end, we need to calculate the sum of all the contributions to these probabilities stemming from the nontrivial submodels. It cannot be greater than the quantum probability for these event, but can be lower since the difference can be compensated by the trivial models. This gives the following consistency condition 
\begin{eqnarray}
\label{conditionGPY}
\Delta_{(0,0,0,1)}(\alpha^2, \gamma)=
p(0,0,0,1)
-\left(\pi-2 \right)\sum\limits_{\text{relevant submodels}}A c_1\nonumber\\
=\frac{\alpha ^2}{2}-\frac{1}{48} (\pi -2) \left(2 \alpha ^8+3 \alpha ^6+6 \alpha ^4 \left(\gamma ^2+2\right)+12 \alpha ^2 \gamma ^2+12 \gamma ^2\right)\geq 0.
\end{eqnarray}
Obviously a simmilar condition could be presented for other single-photon events. However, the one above is the most strict of them all.

Under the assumption $\gamma=\alpha^2$, the condition (\ref{conditionGPY}) simplifies to
\begin{equation}
   \alpha ^2\geq\frac{1}{24} (\pi -2) \alpha ^4 \left(8 \alpha ^4+15 \alpha ^2+24\right).
\end{equation}
It is satisfied for 
approximately $\alpha^2<0.58$.

Finally, notice that the value of $\Delta_{(0,0,0,1)}(\alpha^2, \gamma)$ given by Eq. (\ref{conditionGPY}) decreases with the growth of $\gamma$. Thus, the model definitely works for all $\gamma\leq\alpha^2<0.58$. Of course, this not the full range of the model in the parameter space of $\alpha$ and $\gamma$.

\section*{Appendix E -- Experiment (b) quantum probabilities of four-photon events registered in the GPY scheme}
In this section, we are going to calculate the probability of detecting of detecting $k, l, r, s$ photons respectively in modes $c_1, d_1, c_2, d_2$, in the (b) configuration of the experimental setup outlined in the main text. 
It is given by: 
\begin{eqnarray}
\label{GPY}
P(k, l; r, s)=  e^{-2\alpha^2}(1-\gamma^2)\Big | \sum_{q=0}^{k}  \sum_{p=0}^{r} \sum_{t=0}^{N} 
 {k \choose q} {l \choose q'}  {r \choose p} {s \choose p'}
  \frac{\gamma^t (-1)^{q+p} \alpha_1^{k + l -t} \alpha_2^{r + s - t}t! }{\sqrt{2^{k l r s} } \sqrt{k! l! r! s!}}    \Big |^2,
    \end{eqnarray}
where the upper bound in the last sum,  $N=\text{min}(k+l,r+s)$, is a condition imposed by the expansion of the twin beam state in the Fock basis.  

The table below gives probabilities of the \emph{class 1} events reproduced by the LHV model in the main text. 

$
\begin{array}{c@{\hskip 0.5in}c@{\hskip 0.5in}c}
 \text{Events} & \,\,\,\, \text{Probabilities divided by P(0,0,0,0)}  = e^{-2 \alpha ^2} (1-\gamma^2) 
 \,\,\,\,
    & \{A(\mathbf{n})/P(0,0,0,0),\,c_1(\mathbf{n}),\,c_2(\mathbf{n})\}\\
 \left(
\begin{array}{cccc}
 0 & 1 & 0 & 1 \\
 1 & 0 & 1 & 0 \\
\end{array}
\right) & \frac{1}{4} \left(\alpha^4+2\alpha^2 \gamma  \cos (\theta_1+\theta_2)+\gamma ^2\right) & \left\{\frac{1}{4},\,2,\,1\right\} \\
 \left(
\begin{array}{cccc}
 0 & 1 & 0 & 2 \\
 0 & 2 & 0 & 1 \\
 1 & 0 & 2 & 0 \\
 2 & 0 & 1 & 0 \\
\end{array}
\right) & \frac{1}{16}\alpha^2 \left(\alpha^4+4 \gamma  \left(\alpha^2 \cos (\theta_1+\theta_2)+\gamma \right)\right) & \left\{\frac{a^2}{16},\,4,\,4\right\} \\
 \left(
\begin{array}{cccc}
 0 & 1 & 0 & 3 \\
 0 & 3 & 0 & 1 \\
 1 & 0 & 3 & 0 \\
 3 & 0 & 1 & 0 \\
\end{array}
\right) & \frac{1}{96}\alpha^4 \left(\alpha^4+6\alpha^2 \gamma  \cos (\theta_1+\theta_2)+9 \gamma ^2\right) & \left\{\frac{a^4}{96},\,6,\,9\right\} \\
 \left(
\begin{array}{cccc}
 0 & 1 & 1 & 0 \\
 1 & 0 & 0 & 1 \\
\end{array}
\right) & \frac{1}{4} \left(\alpha^4-2\alpha^2 \gamma  \cos (\theta_1+\theta_2)+\gamma ^2\right) & \left\{\frac{1}{4},\,-2,\,1\right\} \\
 \left(
\begin{array}{cccc}
 0 & 1 & 1 & 2 \\
 1 & 0 & 2 & 1 \\
 1 & 2 & 0 & 1 \\
 2 & 1 & 1 & 0 \\
\end{array}
\right) & \frac{1}{32}\alpha^4 \left(\alpha^4+2\alpha^2 \gamma  \cos (\theta_1+\theta_2)+\gamma ^2\right) & \left\{\frac{a^4}{32},\,2,\,1\right\} \\
 \left(
\begin{array}{cccc}
 0 & 1 & 2 & 0 \\
 0 & 2 & 1 & 0 \\
 1 & 0 & 0 & 2 \\
 2 & 0 & 0 & 1 \\
\end{array}
\right) & \frac{1}{16}\alpha^2 \left(\alpha^4+4 \gamma  \left(\gamma -a^2 \cos (\theta_1+\theta_2)\right)\right) & \left\{\frac{a^2}{16},\,-4,\,4\right\} \\
 \left(
\begin{array}{cccc}
 0 & 1 & 2 & 1 \\
 1 & 0 & 1 & 2 \\
 1 & 2 & 1 & 0 \\
 2 & 1 & 0 & 1 \\
\end{array}
\right) & \frac{1}{32}\alpha^4 \left(\alpha^4-2\alpha^2 \gamma  \cos (\theta_1+\theta_2)+\gamma ^2\right) & \left\{\frac{a^4}{32},\,-2,\,1\right\} \\
 \left(
\begin{array}{cccc}
 0 & 1 & 3 & 0 \\
 0 & 3 & 1 & 0 \\
 1 & 0 & 0 & 3 \\
 3 & 0 & 0 & 1 \\
\end{array}
\right) & \frac{1}{96}\alpha^4 \left(\alpha^4-6\alpha^2 \gamma  \cos (\theta_1+\theta_2)+9 \gamma ^2\right) & \left\{\frac{a^4}{96},\,-6,\,9\right\} \\
\end{array}
$

The LHV model does not reproduce the following probabilities of events that belong to the (2\&2) subspace.\\

$
\begin{array}{c@{\hskip 0.5in}c}
 \text{Events} &  \text{Probabilities divided by P(0,0,0,0)}  =e^{-2 \alpha ^2} (1 - \gamma^2) \\
 \left(
\begin{array}{cccc}
 0 & 2 & 1 & 1 \\
 1 & 1 & 0 & 2 \\
 1 & 1 & 2 & 0 \\
 2 & 0 & 1 & 1 \\
\end{array}
\right) & \frac{1}{32} \left(\alpha^8-4 \alpha^4 \gamma ^2 \cos (2 (\theta_1+\theta_2))+4 \gamma ^4\right) \\
 \left(
\begin{array}{cccc}
 0 & 2 & 0 & 2 \\
 2 & 0 & 2 & 0 \\
\end{array}
\right) & \frac{1}{64} \left(\alpha^8+16 \alpha^4 \gamma ^2+4 \alpha^4 \gamma ^2 \cos (2 (\theta_1+\theta_2))+8 \alpha^2 \gamma  \left(\alpha^4+2 \gamma ^2\right) \cos (\theta_1+\theta_2)+4 \gamma ^4\right) \\
 \left(
\begin{array}{cccc}
 0 & 2 & 2 & 0 \\
 2 & 0 & 0 & 2 \\
\end{array}
\right) & \frac{1}{64} \left(\alpha^8+16 \alpha^4 \gamma ^2+4 \alpha^4 \gamma ^2 \cos (2 (\theta_1+\theta_2))-8 \alpha^2 \gamma  \left(\alpha^4+2 \gamma ^2\right) \cos (\theta_1+\theta_2)+4 \gamma ^4\right) \\
 \left(
\begin{array}{cccc}
 1 & 1 & 1 & 1 \\
\end{array}
\right) & \frac{1}{16} \left(\alpha^8+4 \alpha^4 \gamma ^2 \cos (2 (\theta_1+\theta_2))+4 \gamma ^4\right)
\end{array}
$


\section*{Appendix F -- A simple calculation of CH inequality violation by on/off version of the TWC experiment (a) for very weak local oscillators}
In this section we show the {\em method} of the derivation of the probabilities appearing in Eq. (10) of the main text. 
We present this for an approximation in which the coherent local oscillator fields are replaced by their first two terms.
Thus what we present here is just illustrative. We want to avoid unnecessary technicalities.

For simplicity we take the single photon state as $\frac{1}{\sqrt{2}}(b_1^\dagger+b_2^\dagger)$.
We have the following initial state:
\begin{eqnarray}
\label{TWC_app}
    \frac{1}{\sqrt{2}}( b_1^\dagger+b_2^\dagger)\frac{1}{1+\alpha^2}
    (1+\alpha_1a^\dagger_1)
    (1+\alpha_2a^\dagger_2)|\Omega\rangle. 
\end{eqnarray}
When the state in (\ref{TWC_app}) impinges on the beamsplitters of transitivity $T$, and reflectively $R$, 
it transforms to 
\begin{eqnarray}
    \frac{1}{\sqrt{2}}
    \left( \sqrt{T}d_1^\dagger+i\sqrt{R}c_1^\dagger+\sqrt{T}d_2^\dagger+ i\sqrt{R}c_2^\dagger\right)
    \frac{1}{1+\alpha^2}
    \left[1+\alpha_1(\sqrt{T}c_1^\dagger+ i\sqrt{R}d_1^\dagger)\right]
    \left[1+\alpha_2(\sqrt{T}c_2^\dagger+ i\sqrt{R}d_2^\dagger))\right]|\Omega\rangle.
\end{eqnarray}
The probability $P(A',B')$ is related to the amplitude of detecting 
a photon in both detectors $D_{d_1}$ and $D_{d_2}$, namely to the amplitude $\frac{1}{\sqrt{2} (1+\alpha^2)}     \left(\sqrt{T}i\sqrt{R}\alpha_1 + \sqrt{T}i\sqrt{R}\alpha_2 \right)$.
That gives the probability $P(A',B')=\frac{T R }{2}
    \frac{\left|\alpha_1+\alpha_2\right|^2}{(1+\alpha^2)^2}.$
    
The event A  (B) is defined as the firing of any any local detector when the local oscillator is off, that gives the trivial probabilities $P(A)=1/2=P(B)$ and $P(A,B)=0. $
While, the probability of the event pair $P(A,B')$ is related to the final state  
    \begin{eqnarray}
    \frac{1}{\sqrt{2}} \left(b_1^\dagger+\sqrt{T}d_2^\dagger+ i\sqrt{R}c_2^\dagger\right)
    \frac{1}{\sqrt{1+\alpha^2}}
    \left[1+\alpha_2(\sqrt{T}c_2^\dagger+ i\sqrt{R}d_2^\dagger)\right]|\Omega\rangle
\end{eqnarray}
The amplitude of $b^\dagger_1d_2^\dagger|\Omega\rangle$ is
$\frac{i}{\sqrt{2}}
    \frac{\sqrt{R}\alpha_2}{\sqrt{1+\alpha^2}}$. 
    Thus, $P(A,B')=  \frac{{R}\alpha^2}{2(1+\alpha^2)} $ and so is $P(A',B)$.
    
    We take the left hand side CH inequality
    \begin{eqnarray}
    \label{CHin2}
        -1\leq P(A,B)+P(A,B')+P(A',B)-P(A',B')
        -P(A)-P(B) = CH \leq 0, ~~~
    \end{eqnarray}
    and put the values of the probabilities, results:
    $ 0+\frac{\alpha^2}{{1+\alpha^2}}R-\frac{1}{{2}}
    {T}{R}\left|\alpha_1+\alpha_2\right|^2
    \frac{1}{(1+\alpha^2)^2}
    -1/2-1/2$. 
    We choose the phases of the coherent states to be identical to get
    $ \frac{\alpha^2}{{1+\alpha^2}}R-2
    {T}{R}\alpha^2
    \frac{1}{(1+\alpha^2)^2}
    -1= \frac{\alpha^2}{{1+\alpha^2}}R\left(1-2T  \frac{1}{1+\alpha^2}\right)-1$, which obviously can be less that $-1$, with a proper choice of $T$.  The CH inequality will be  violated when 
    $T> \frac{1+\alpha^2}{2} > \frac 12$. 
    
    Note that the situation with balanced beamsplitters, $T=1/2,$ does not violate the CH inequality.
    
    
    
    
    
\end{document}